\documentstyle[aas2pp4]{article}
\lefthead{Balogh et al.}
\righthead{Star Formation in Cluster Galaxies at $0.2<z<0.55$}

\begin{document}
\newcommand{\oii}{[OII]$\lambda$3727}
\newcommand{\ow}{$W_{\circ}(OII)$}
\title{Star Formation in Cluster Galaxies at $0.2<z<0.55$}

\author{Mike L. Balogh\altaffilmark{1}, Simon L. Morris\altaffilmark{2, 5}, H. K. C. Yee\altaffilmark{3, 5},}
\author{R.G. Carlberg\altaffilmark{3, 5}, and Erica Ellingson\altaffilmark{4, 5}}
\altaffiltext{1}{\small{Department of Physics \& Astronomy, University of Victoria, Victoria, BC, V8X 4M6, Canada. \\ email: balogh@uvphys.phys.uvic.ca}}

\altaffiltext{2}{\small{Dominion Astrophysical Observatory, National Research Council, 5071 West Saanich Road, Victoria, B.C., V8X 4M6 Canada. email: Simon.Morris@hia.nrc.ca}}

\altaffiltext{3}{\small{Department of Astronomy, University of Toronto, Toronto, Ontario, M5S 1A7 Canada. \\ email: hyee, carlberg@astro.utoronto.ca}}

\altaffiltext{4}{\small{CASA, University of Colorado, Boulder, Colorado 80309-0389. \\ email: e.elling@pisco.colorado.edu}}
\altaffiltext{5}{\small{Visiting Astronomer, Canada--France--Hawaii Telescope, which is operated by the National Research Council of Canada, le Centre National de Recherche Scientifique, and the University of Hawaii.}}

\begin{abstract}
The rest frame equivalent width of the \oii\ emission line, \ow, has been measured for cluster and field galaxies in the CNOC redshift survey of rich clusters at $0.2<z<0.55$.  Emission lines of any strength in cluster galaxies at all distances from the cluster centre, out to $2R_{200}$, are less common than in field galaxies.  The mean \ow\ in cluster galaxies more luminous than $M_r^k<-18.5 + 5\log{h}$ (q$_{\circ}$=0.1) is $3.8 \pm 0.3$ \AA\ (where the uncertainty is the 1$\sigma$ error in the mean), significantly less than the field galaxy mean of $11.2 \pm 0.3$ \AA.  For the innermost cluster members ($R<0.3R_{200}$), the mean \ow\ is only $0.3 \pm 0.4$ \AA.  Thus, it appears that neither the infall process nor internal tides in the cluster induce detectable excess star formation in cluster galaxies relative to the field.  The colour--radius relation of the sample is unable to fully account for the lack of cluster galaxies with $W_{\circ}(OII)>10$ \AA, as expected in a model of cluster formation in which star formation is truncated upon infall.  Evidence of supressed star formation relative to the field is present in the whole cluster sample, out to 2$R_{200}$, so the mechanism responsible for the differential evolution must be acting at a large distance from the cluster centre, and not just in the core.  The mean star formation rate in the cluster galaxies with the strongest emission corresponds to an increase in the total stellar mass of less than about 4\% if the star formation is due to a secondary burst lasting 0.1 Gyr.
\end{abstract}


\section{Introduction} \label{sec-intro}
It is well established that galaxy populations vary with the density of neighbouring galaxies (e.g., \cite{Dress}; \cite{WGJ}); however, the physical mechanisms responsible for the variation are not known.  It has also been observed that cluster galaxies have, on average, older stellar populations than field galaxies (e.g., \cite{B+90}; \cite{R+94}).  Thus, if clusters evolve by accreting field galaxies, star formation in the infalling galaxies must be truncated prematurely, relative to isolated field galaxies.  If clusters are to be used to determine the mass density of the universe (e.g., \cite{C+96}), the effect of this differential evolution between cluster and field galaxies on the average galaxy stellar mass must be understood.

Star formation may be truncated following an increase in star forming activity which rapidly consumes and/or expels the available gas in a galaxy.   Several physical processes have been proposed which may have such an effect, including shocks induced by ram pressure from the intracluster medium (ICM, \cite{BD86}; \cite{GJ}), effects of the cluster tidal field (\cite{BV}), and galaxy--galaxy interactions (\cite{BH}; \cite{M+96}).  The increase in the fraction of blue, star forming cluster galaxies with redshift (BO effect, \cite{BO}), has been well established, and several authors (e.g., \cite{CS}, \cite{MW}; \cite{Caldwell}; \cite{Barger}) have shown that there are cluster galaxies, even at low redshift, in which significant star formation has occurred in the last 2 Gyr. It is not yet clear, however,  whether or not this activity is in excess relative to the field. 

Alternatively, star formation may be halted in infalling galaxies without an initial increase, as suggested by the results of the analysis of colours, spectral features and morphologies of galaxies in the Abell 2390 cluster (\cite{A+96}).  This may be achieved by interaction with the hot ICM by ram pressure stripping (\cite{GG}) or transport processes such as viscous stripping and thermal evaporation (\cite{N82}).  In this case, cluster galaxies can be treated as representative of the field at the epoch of infall, and the BO effect is interpreted as an increase in the infall rate of field galaxies, which themselves show evidence of more star forming activity at higher redshift.

The luminosities of Balmer emission lines in galaxy spectra are directly related to the ionising fluxes of hot stars embedded in HII regions, and thus can be used to determine the star formation rate (SFR) in the observed region of the galaxy (\cite{K92}).  Although H$\alpha$ is the best observable indicator of SFR, it is redshifted out of convenient observing bands at even moderate redshifts.  The \oii\ emission line is then the feature of choice, as its strength is found to be correlated with H$\alpha$ in local samples (\cite{K92}, \cite{Guzman}, but see \cite{CFRSXIV}).  It has been clearly shown (e.g., Dressler, Thompson \& Shectman 1985; \cite{HO}; \cite{A+96}; \cite{B+97}) that the fraction of galaxies with strong emission lines is much smaller in clusters than in the field.   Since emission lines are much more commonly found in late spirals than in early type galaxies (e.g., \cite{K92}; \cite{B+97}), this effect may be consistent with the morphology--radius relation, if the fraction of spiral galaxies  is lower in clusters by the amount necessary to account for the decrease in observed emission.  However, if star formation is truncated in field galaxies falling into the cluster, the number of galaxies with [OII] line emission will be lower than expected from the morphological composition at a given cluster--centric radius, as the [OII] feature disappears shortly after star formation ceases, whereas morphological change due to disk fading occurs on timescales of about 1 Gyr (\cite{A+96}).
 
 In this work, the dependence of [OII] line strength on distance from the cluster centre is presented and compared with the field sample.   In Section \ref{sec-sample} the data sample is described, selection effects are considered, and cluster membership and cluster--centric radius are defined.  In Section \ref{sec-res} the emission line properties of cluster galaxies are compared with the field sample.  The results are interpreted in Section \ref{sec-discuss} by computing star formation rates and comparing the fraction of emission line galaxies with the colour--radius relation. The conclusions are summarized in Section \ref{sec-sum}.  Throughout this {\em Letter}, a cosmology of $q_{\circ}=0.1$ is assumed for distance dependent calculations, which are given in terms of $h=H_{\circ}/100$. 

\section{Sample Selection and Measurements} \label{sec-sample}
The galaxy sample was selected from the Canadian Network for Observational Cosmology (CNOC, Yee, Ellingson \& Carlberg 1996, \cite{YEC}) spectroscopic sample of fifteen\footnote{Omitting cluster E0906+11, for which a velocity dispersion could not be computed (\cite{C+96}).} rich, X--ray luminous clusters at moderate  redshift ($0.2<z<0.55$).  This sample consists of about 2500 cluster and field galaxies with determined redshifts, for which selection effects are well understood (\cite{YEC}). 

For each spectrum the rest frame \oii\  equivalent width, \ow, was automatically computed by summing the flux above the continuum in pixels between $3713<\lambda< 3741$ \AA.  The continuum level was estimated by fitting a straight line to the flux between $3653 <\lambda < 3713$ \AA\ and $3741<\lambda < 3801 $ \AA\ using weighted linear regression, with weights from the Poisson noise vector generated by optimally extracting the spectra with IRAF\footnote{IRAF is distributed by the National Optical Astronomy Observatories which is operated by AURA Inc. under contract with NSF.}.  The error in \ow\ is computed from equation A8 in Bohlin et al. (1983).   An average \ow, weighted by this error, is adopted for multiply observed galaxies in the sample.  The mean and median error in \ow\ is 5 \AA\ and 3 \AA, respectively, for the full sample.  The accuracy of the  measurements and errors was verified by comparing measurements made on artificial spectra which consist of a power law continuum component ($f_\nu \propto \nu^{0.5}$) added to the spectrum of M31 (making the bulge spectrum mimic a late type spiral in the continuum), and a Gaussian emission line at $\lambda=3727$ \AA\ with a velocity width of 5\AA\ (400 km/s) FWHM.   The standard deviation of \ow\ measurements for each set of 250-1000 spectra at the same signal-to-noise ratio (SNR) and \ow\ was found to compare well with the average error estimate. In addition, the difference between two independent measurements of the same (real) galaxy, when available, was compared with $\sigma$, the quadrature sum of the two error estimates.  This analysis indicates that the \ow\ errors do not represent a normal distribution, as only about 50\% of the differences between two measurements are less than 1$\sigma$, and only 92\% are less than 3$\sigma$.  The quoted error estimates are still meaningful, however, so long as they are interpreted in this sense.  A copy of the FORTRAN code used to measure \ow\ and its error (as well as several other indices) can be obtained from the first author.

The CNOC selection procedure is described in YEC, and is designed to sample the cluster galaxies  to $M_r \lesssim -18.5 + 5\log{h}$ with at least an 80\% success rate. A magnitude weight $W_m$, which is the ratio of the total number of galaxies to the number of galaxies with redshifts in a magnitude bin centred around the galaxy, is calculated for each galaxy in the sample to correct for incompleteness.  To ensure that the sample is not biased toward emission line objects, galaxies with $W_m>5$ are excluded.  The remaining, magnitude weighted sample is complete to about $M_r^k = -18.5 + 5\log{h}$; galaxies less luminous than this limit are excluded from the sample.  

Cluster velocity dispersion profiles of the form $\sigma^2(r)=B/(r+b)$, where r is the projected radius from the cluster centre, are calculated by Carlberg, Yee \& Ellingson (1997), based on a volume density function of the form $\nu(r)=A/r^{-1}(r+a)^{-3}$ and an anisotropy parameter $\beta=0.5$.  Galaxies with a velocity difference relative to the cluster mean of less than $3\sigma (r)$ are considered to be cluster members.  The field sample is selected from galaxies with a velocity difference greater than $6\sigma(r)$, and which lie within a filter dependent redshift range which minimizes selection effects (YEC).  The population with intermediate velocities is classified as ``near--field'', and may contain infalling field galaxies.  These galaxies are not included in the present analysis to ensure as clear a differentiation between field and cluster galaxies as possible.  The \ow\ properties of this population are not, however, statistically different from those of the field.

The cluster--centric distance R for cluster members is defined as the projected distance from the brightest cluster galaxy (BCG). For field galaxies, R is the redshift difference from the cluster average assuming Hubble flow.  Since the sample consists of clusters of different richness, R is normalised by $R_{200}$, the radius at which the cluster mass density is 200 times the critical density\footnote{The overdensity of virialisation is approximately $\delta \rho / \rho = 178\Omega^{-0.6}$, which corresponds to a radius of about $1.5 R_{200}$ for $\Omega=0.2$.}.  For these clusters, $R_{200}$ is typically 1--1.5 h$^{-1}$ Mpc (\cite{C+96}).  There is an apparent absence of field galaxies in the sample at $3<R/R_{200}<20$, due to the fact that field galaxies at that redshift, projected in front of and behind the cluster, have a velocity offset from the BCG less than 3$\sigma(r)$ and are hence included in the cluster sample.  Limited spatial coverage on the sky restricts the observed projected distance of galaxies in the sample to less than about 3R$_{200}$.  For statistical analysis, each galaxy is weighted by $W_m*W_{ring}$, where $W_{ring}$ is a geometrical correction to account for the fact that the clusters are not uniformly sampled as a function of radius.

The restricted sample considered in this analysis consists of 727 cluster galaxies and 346 field galaxies, whereas \ow\ measurements are available for a total of 1169 cluster and 783 field galaxies.   The BCGs are considered atypical cluster members, and are excluded from all analysis.  Also, no attempt was made to identify active galactic nuclei (AGN), as the H$\beta$ and [OIII]$\lambda$5007 lines, which are common diagnostics, are usually redshifted out of the observed spectral range.  

\section{Results}\label{sec-res}
Figure \ref{fig-oiirad} shows the distribution of \ow\ as a function of R/R$_{200}$, where R is defined in Section \ref{sec-sample}.  The dashed line separates the (inner) cluster galaxies from the field galaxies.   As expected, emission line galaxies are clearly less common in clusters than in the field.  The weighted mean \ow\ (and 1$\sigma$ uncertainty) in the field is $11.2 \pm 0.3$ \AA, compared with $3.5 \pm 0.4$ \AA\ in the outer cluster regions ($0.3<R/R_{200}<2$) and $0.3 \pm 0.4$ \AA\ in the central regions ($R/R_{200}<0.3$).  The mean error of an individual measurement, indicated by the sample error bar in the Figure, is 3.5 \AA.  (Forty of the 1073 galaxies in the selected sample have (formal) uncertainties in \ow\ greater than 10 \AA\, and 117 have uncertainties less than 1 \AA).  There is no evidence of a population of cluster galaxies with excess emission relative to the field at any distance from the cluster centre, out to $R\approx2R_{200}$.  

The difference between the cumulative \ow\ distributions of the cluster and field is shown in the top panel of Figure \ref{fig-oiisfrcum}.  The cluster galaxies at $R \ge 0.3R_{200}$ are represented by the solid line, the inner cluster galaxies ($R<0.3R_{200}$) by the long dashed line and the field galaxies by the dotted line.  The cluster sample shows a clear deficiency in emission line galaxies relative to the field in both the inner and outer cluster regions at all line strengths.  There is no evidence of a population of cluster galaxies with stronger \ow\ than is observed in field galaxies.  Since the cluster sample is partially contaminated by field galaxies projected on the cluster, the measurements of the mean \ow\ and SFR in the cluster are overestimates.

\section{Discussion} \label{sec-discuss}

Star formation rates (SFRs) have been calculated from Kennicutt's (1992) relation with his adopted extinction correction of E(H$\alpha$) = 1 mag:
\begin{equation}
\label{eqn:sfr}
SFR(M_{\sun}yr^{-1})=6.75 \times 10^{-12} {L_B \over L_B(\sun)}W_{\circ}(OII),
\end{equation}
where $L_B/L_B(\sun)=10^{0.4(5.48-M_B)}$ depends on the absolute $B$ band luminosity of the galaxy, which must be obtained from the available Gunn $g$ and $r$ photometry: $M_B=M_r+(g-r)_{\circ}-(g-B)_{\circ}$.  Rest frame $(g-r)_{\circ}$ colours are computed from the colour--redshift relations in Patton et al. (1997; their Figure 7), which are fits to the colour k--corrections of YEC, and the corresponding rest frame $(g-B)_{\circ}$ colour is found by linearly interpolating the published values in Fukugita, Shimasaku \& Ichikawa (1995; their Table 3f).  
The cumulative SFR distributions for the cluster and field  populations are shown in the bottom panel of Figure \ref{fig-oiisfrcum}.    
For the inner cluster members, less than 35\% have a SFR$>$0.01 $h^{-2} M_{\sun} \mbox{yr}^{-1}$, whereas the median SFR in the field is about 0.2 $h^{-2} M_{\sun} \mbox{yr}^{-1}$.  The SFR calculated in this manner is most useful as an indication of the relative difference between the cluster and field; Guzman et al. (1997) suggest that the coefficient in equation \ref{eqn:sfr} may be about three times lower than is used here.  

It has been clearly shown (e.g., \cite{CS}, \cite{MW}; \cite{Caldwell}; \cite{Barger}) that a significant fraction of cluster galaxies have undergone episodes of star formation in the last 2 Gyr.  In particular, Barger et al. (1996) suggest that 30\% have undergone a 0.1 Gyr burst in the last 2 Gyr, which implies that 1.5\% of cluster galaxies should be in such a state at any one time.  There are cluster galaxies in the present sample with non-zero \ow; however, they are less common than in the field population. For example, 4.3\% of field galaxies have $W_{\circ}(OII)>40$ \AA, compared with only 1.4\% of cluster galaxies. Thus, it seems unlikely that significant additional star formation activity in cluster galaxies is caused by the infall process or internal tides in the cluster.  The weighted mean \ow\ of the cluster galaxies with $W_{\circ}(OII)>40$ \AA\ is 59 \AA, which corresponds to an increase in stellar mass of only 4\% over 0.1 Gyr from equation \ref{eqn:sfr}, assuming a stellar mass-to-light ratio of unity.   No galaxy anywhere in the sample is observed to have  $W_{\circ}(OII)>150$ \AA, or $SFR > 4M_{\sun}yr^{-1}$; if there are galaxies with $SFR\approx 30 M_{\sun}yr^{-1}$ as suggested by Couch \& Sharples (1987) and Barger et al. (1996), they may be located outside the sample, at $R>2R_{200}$.

It is instructive to compare the \ow--radius relation with the well known morphology--radius relation, to determine whether or not they are consistent with one another.   Unfortunately, morphological classifications are not yet available for the full CNOC sample.  For now,  the colour--radius relation is considered, as galaxy morphology is expected to be correlated with colour.  The  rest frame $(g-r)_{\circ}$ colours, computed as described above, are used to divide the sample into four classes, which correspond roughly to E, Sbc, Scd and Im morphological types, as in Patton et al. (1997).   The fraction of field galaxies with $W_{\circ}(OII)>10$ \AA\ is 0.12$\pm0.04$, 0.29$\pm0.06$, 0.72$\pm0.13$ and 0.78$\pm0.17$  for the E, Sbc, Scd and Im classes, respectively.    From the colour--radius relation of the cluster sample, the fraction of cluster galaxies with \ow $>10$ \AA\ in a given radial bin is predicted; this is shown as the dashed line in Figure \ref{fig-oiimtype}.  The observed fraction is shown as the solid line; it is significantly lower than expected from the colour--radius relation alone, for $R<2R_{200}$.  The fraction of galaxies with \ow $>15$ \AA, however, is consistent with the colour--radius relation; this may suggest that star formation is not truncated equally for all galaxy types, as already suggested by the results of Moss \& Whittle (1993).  

\section{Conclusions}
\label{sec-sum}
The  mean \ow\ of cluster galaxies more luminous than $M_r^k<-18.5 + 5\log{h}$ in the CNOC spectroscopic sample of rich clusters at $0.2<z<0.55$ is $3.8 \pm 0.3$ \AA, significantly less than the field galaxy mean of $11.2 \pm 0.3$ \AA.  The average SFR among cluster galaxies is less than the average in the field out to 2$R_{200}$,  which implies that whatever mechanism is responsible for truncating star formation in cluster galaxies is taking place at a large distance from the cluster centre.   Cluster galaxies of a given colour are less likely to show signs of significant star formation than their counterparts in the field at any distance from the cluster centre.  Of cluster members, 1.4\% have $W_{\circ}(OII)>40$ \AA, with a weighted mean of 59 \AA, corresponding to an increase in stellar mass  of less than 4\% if the activity is due to a 0.1 Gyr burst.   Many more (4.3\%) field galaxies have $W_{\circ}(OII)>40$ \AA, suggesting that star formation in cluster galaxies is likely not induced by the infall process or internal tides.  This supports the conclusions of Abraham et al. (1996) that star formation is truncated in infalling field galaxies without an initial increase.  The BO effect in these clusters may then be due to the increased rate of infall of bluer field galaxies at higher redshift.

\acknowledgments
We would like to thank the referee, J. A. Rose, for improving the clarity and focus of this work.  MLB would also like to thank C. J. Pritchet for helpful comments and discussions.  MLB is supported by the Natural Sciences and Engineering Research Council of Canada.

\clearpage
\begin{figure}
\caption{\ow\ as a function of cluster--centric distance R, for the selected subsample.  The points to the left of the dashed line are cluster galaxies, for which R is the projected distance from the cluster centre.  There is one cluster galaxy, with \ow=150 \AA, which is off the scale.  The more distant points are field galaxies, for which R is the Hubble flow distance determined from the redshift difference between the galaxy and the cluster mean.  The solid line is the weighted mean in the field and three cluster radial bins.  The sample error bar displayed is representative of the mean 1$\sigma$ uncertainty in \ow, 3.5 \AA.\label{fig-oiirad}}
\end{figure}


\begin{figure}
\caption{Top panel: The cumulative distribution of \ow\ in the inner cluster ($R/R_{200}<0.3$, long dashed line), outer cluster ($0.3<R/R_{200}<2$, solid line) and field (dotted line) populations.  Bottom panel: The cumulative distribution of star formation rates for the same three populations, calculated from the \ow\ and Kennicutt's (1992) relation as described in the text.  Note that galaxies with negative \ow, produced by random errors about zero, correspond to negative SFRs; thus, the cumulative functions do not reach unity on this plot.\label{fig-oiisfrcum}}
\end{figure}

\begin{figure}
\caption{The fraction of cluster galaxies with \ow $>$ 10 \AA\ as a function of normalised distance from the cluster centre (solid line), compared with the fraction that would be expected from the colour--radius relation (long--dashed line).  The horizontal, short--dashed line is the fraction in the field galaxy sample.  See Section \ref{sec-discuss} for details.  \label{fig-oiimtype}}
\end{figure} 


\clearpage

\begin{thebibliography}{}

\bibitem[Abraham et al. 1996]{A+96} Abraham, R. G. et al., \apj, 471, 694
\bibitem[Barger et al. 1996]{Barger}Barger, A. J., Aragon-Salamanca, A., Ellis, R. S., Couch, W. J., Smail, I., \& Sharples, R. M. 1996, \mnras, 279, 1
\bibitem[Barnes \& Hernquist 1991]{BH}Barnes, J. E., \& Hernquist, L. E. 1991, \apj, 370, L65
\bibitem[Biviano et al. 1997]{B+97} Biviano, A., Katgert, P., Mazure, A., Moles, M., den Hartog, R., Perea, J., \& Focardi, P. 1997, \aap, 321, 84
\bibitem[Bohlin et al. 1983]{Bohlin} Bohlin, R. C., Hill, J. K., Jenkins, E. B., Savage, B. D., Snow,T. P. Jr., Spitzer, L. Jr., \& York, D. G. 1983, \apjs, 51, 277
\bibitem[Bower et al. 1990]{B+90}Bower, R. G., Ellis, R. S., Rose, J. A., \& Sharples, R. M. 1990, \aj, 99, 2
\bibitem[Bothun \& Dressler 1986]{BD86}Bothun, G. D., \& Dressler, A. 1986, \apj, 301, 57
\bibitem[Butcher \& Oemler 1984]{BO}Butcher, H., \& Oemler, A. 1984, \apj, 285, 426
\bibitem[Byrd \& Valtonen 1990]{BV}Byrd, G., \& Valtonen, M. 1990, \apj, 350, 89
\bibitem[Caldwell et al. 1996]{Caldwell}Caldwell, N., Rose, J. A., Franx, M. \& Leonardi, A. J. 1996, \aj, 111, 78
\bibitem[Carlberg et al. 1996]{C+96} Carlberg, R. G., Yee, H. K. C., Ellingson, E., Abraham, R., Gravel, P., Morris, S., \& Pritchet, C. J. 1996, \apj, 462, 32
\bibitem[Carlberg, Yee \& Ellingson 1997]{CYE}Carlberg, R. G., Yee, H. K. C., \& Ellingson, E. 1997, \apj, 478, 462
\bibitem[Couch \& Sharples 1987]{CS}Couch, W. J., \& Sharples, R. M. 1987, \mnras, 229, 423
\bibitem[Dressler 1980]{Dress}Dressler, A. 1980, \apj, 236, 351
\bibitem[Dressler et al. 1985]{D+85} Dressler, A., Thompson, I. B., \& Shectman, S. 1985, \apj, 288, 481
\bibitem[Fukugita et al. 1995]{F+95} Fukugita, M., Shimasaku, K., \& Ichikawa, T. 1995, \pasp, 107, 945
\bibitem[Gavazzi \& Jaffe 1987]{GJ}Gavazzi, G., \& Jaffe, W. 1987, \apj, 310, 53
\bibitem[Gott \& Gunn 1972]{GG} Gott, J. R., \& Gunn, J. 1972, \apj, 176, 1
\bibitem[Guzman et al. 1997]{Guzman}Guzman, R., Gallego, J., Koo, D. C., Phillips, A. C., Lowenthal, J. D., Faber, S. M., Illingworth, G. D., \& Vogt, N. P. 1997, astro-ph/9704001
\bibitem[Hammer et al. 1997]{CFRSXIV}Hammer et al. 1997, \apj, 481, 49
\bibitem[Hernquist, L. 1990]{Hernquist}Hernquist, L. 1990, \apj, 356, 359
\bibitem[Hill \& Oegerle 1993]{HO} Hill, J. M., \& Oegerle, W. R. 1993, \aj, 106, 3
\bibitem[Kennicutt 1992]{K92}Kennicutt, R. C., Jr. 1992, \apj, 388, 310
\bibitem[Moore et al. 1996]{M+96}Moore, B., Katz, N., Lake, G., Dressler, A. \& Oemler, A. 1996,  \nat, 379, 613
\bibitem[Moss \& Whittle 1993]{MW}Moss, C., \& Whittle, M. 1993, \apj, 407, L17
\bibitem[Nulsen 1982]{N82}Nulsen, P. E. J. 1982, \mnras, 198, 1007
\bibitem[Patton et al. 1997]{P+97}Patton, D. R., Pritchet, C. J., Yee, H. K. C., Elingson, E., \& Carlberg, R. G. 1997, \apj, 475, 29
\bibitem[Rose et al. 1994]{R+94}Rose, J. A., Bower, R. G., Caldwell, N., Ellis, R. S., Sharples, R. M., \& Teague, P. 1994, \aj, 108, 6
\bibitem[Whitmore, Gilmore \& Jones 1993]{WGJ} Whitmore, B. C., Gilmore, D. M., \& Jones, C. 1993, \apj, 407, 489
\bibitem[YEC]{YEC} Yee, H. K. C., Ellingson, E., \& Carlberg, R. G. 1996, \apjs, 102, 629 (YEC)
\end{thebibliography}
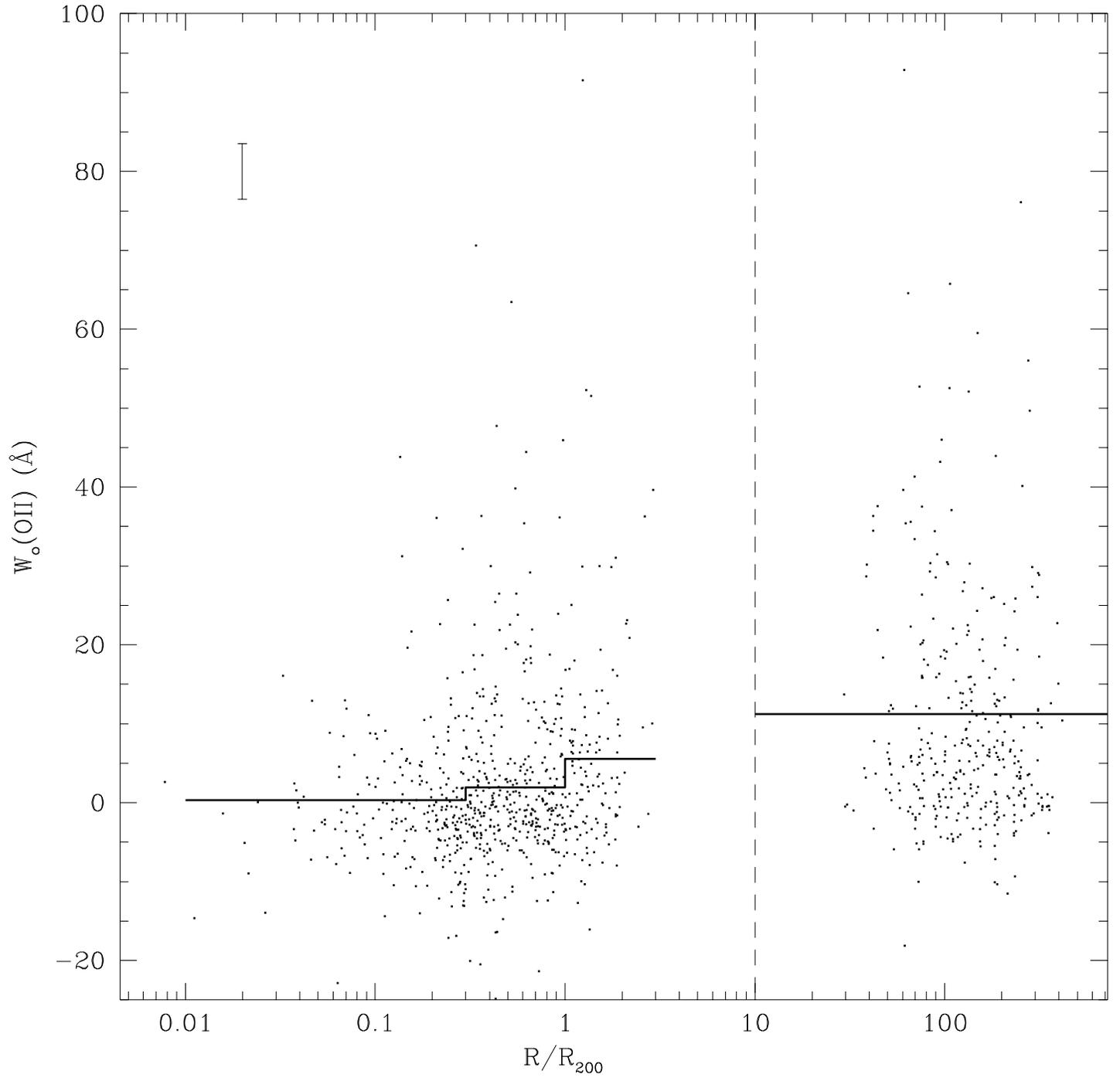
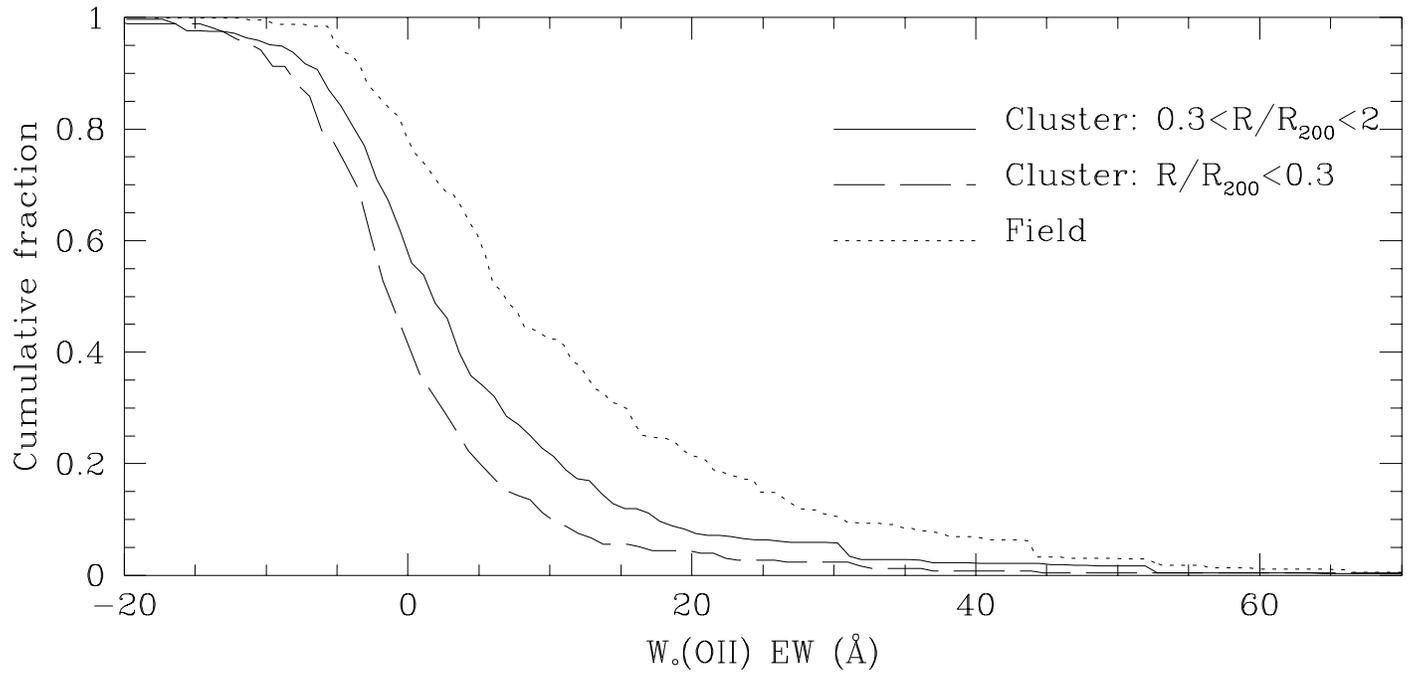
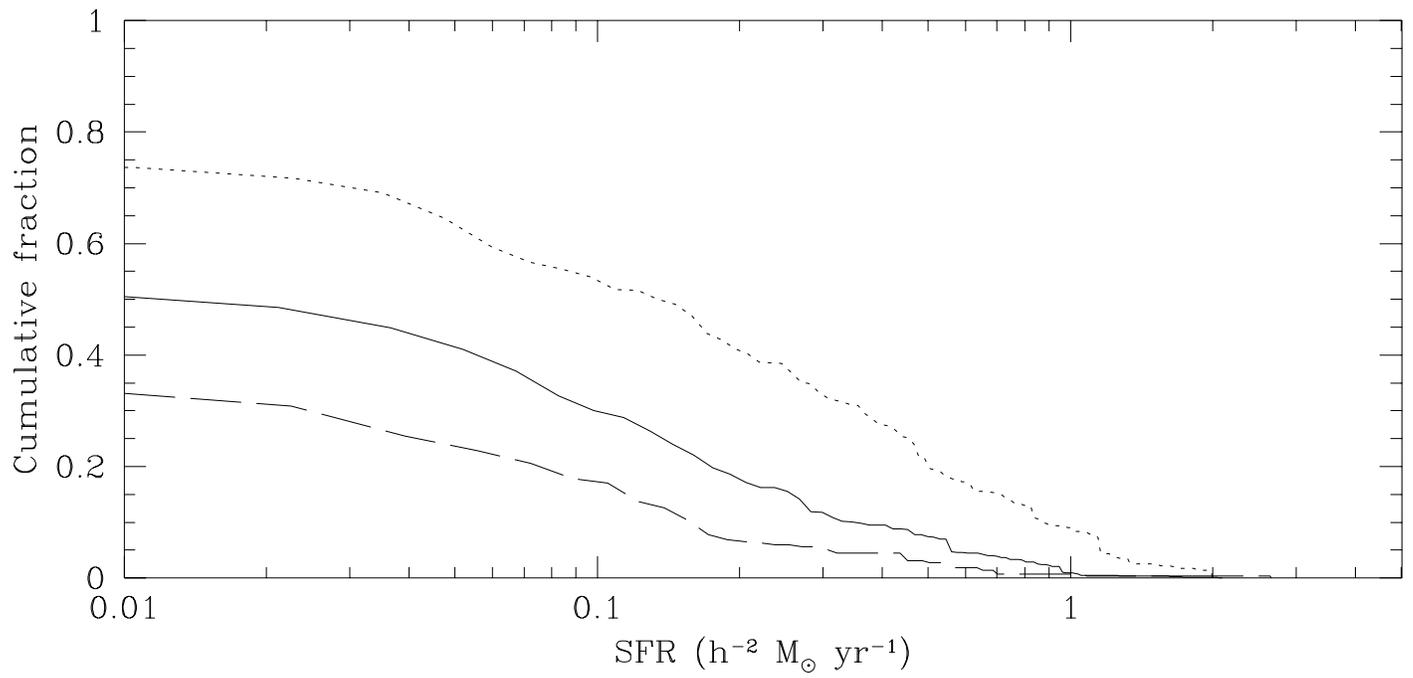
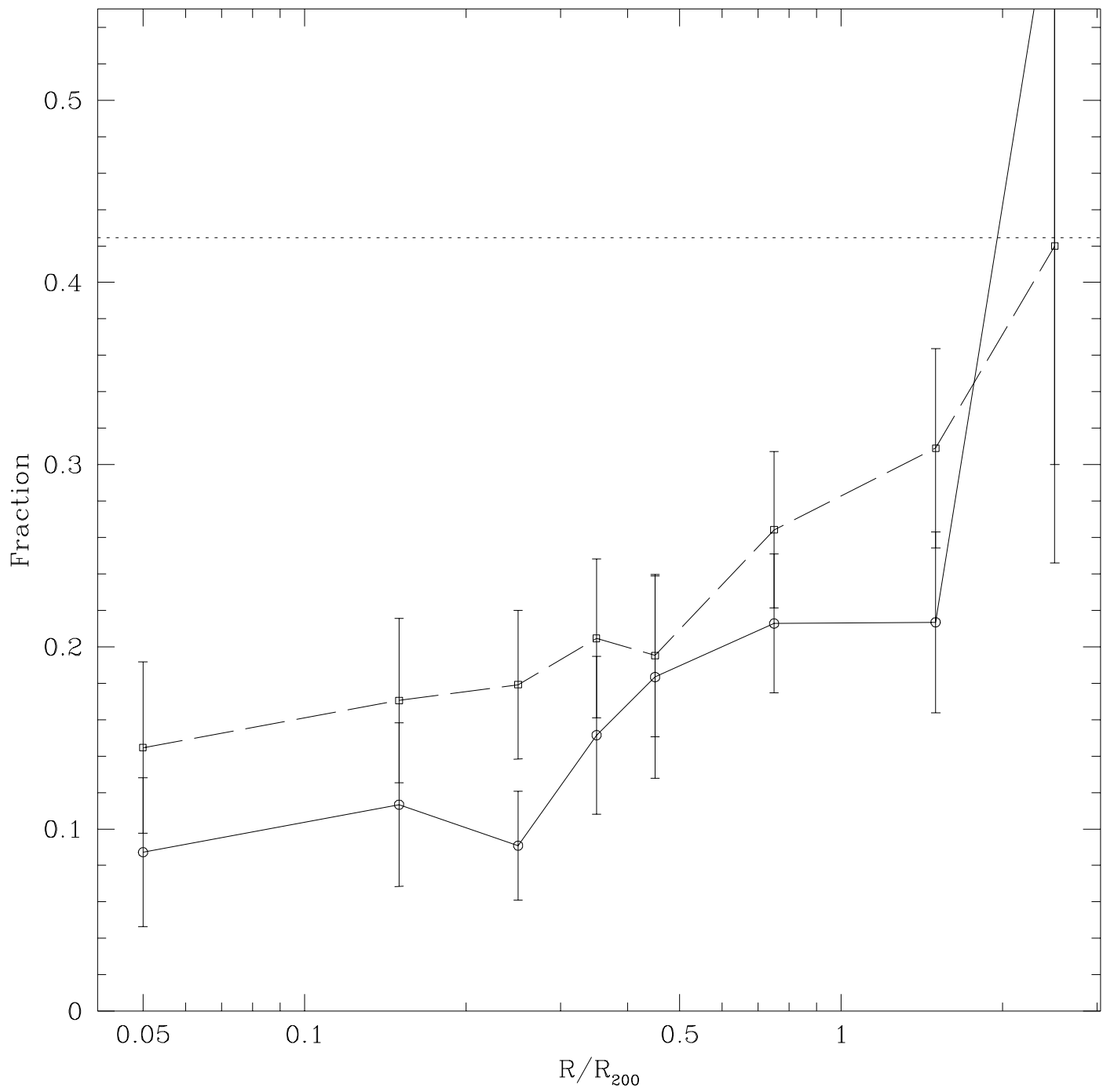
\end{document}